\documentclass[a4paper,11pt]{article}
\usepackage{pos}
\usepackage{graphicx}
\usepackage{float}

\title{Study of solar activity with AERA data}
\ShortTitle{Study of solar activity with AERA data}

\author[a]{D. C dos Santos}

\author[b, \dag]{ for the Pierre Auger Collaboration\notes{\note{Full 
author list at \url{http://www.auger.org/archive/authors_2024_06.html}.}}}

\affiliation[a]{Instituto de Física, Universidade Federal do Rio de Janeiro,\\
  Rio de Janeiro, Caixa postal 68528, Rio de Janeiro, Brazil}

\affiliation[b]{Observatorio Pierre Auger, Av. San Mart\'in Norte 304,
5613 Malarg\"ue, Argentina.}

\emailAdd{spokespersons@auger.org}

\abstract{We investigate the effects of solar activity on the Auger Engineering Radio Array  (AERA)  data collected over 10 years. We report the observation of Solar Radio Burst signals in AERA data associated with intense solar flares accompanied by moderate and strong radio blackout levels recorded by the National Oceanic and Atmospheric Administration (NOAA). Additionally, although in a frequency range different from AERA,  the increased level of X-ray and extreme ultraviolet radiation during periods of larger solar activity also impacts the AERA data in a twofold way: (i) causing the atmosphere to bounce terrestrial radio waves emitted by sources far from the AERA site or (ii) causing radio signals to become degraded or completely absorbed (radio blackout) if the solar radiation results in larger ionization of the upper or lower layers of the ionosphere, respectively. We describe the identification of both cases in AERA data with a remarkable correlation between the Maximum Usable Frequency (MUF), that represents the highest frequency that can be used for radio communication between two points located on the Earth, modulated by the solar cycle, and the broadband noise observed  in the frequency range of 30-40 MHz.}

\FullConference{%
  10th International Workshop on Acoustic and Radio EeV Neutrino Detection Activities - ARENA2024\\
  11-14 June 2024\\
  The Kavli Institute for Cosmological Physics, Chicago, IL, USA}

\begin{document}
\maketitle

\section{Introduction}
\label{sec:intro}

\hspace{0.5cm}The Auger Engineering Radio Array (AERA)\cite{AERA}, part of the Pierre Auger Observatory, located at Malargüe, Argentina, is a facility designed to detect radio emissions from extensive air showers produced by high-energy cosmic rays. It consists of 153 autonomous radio-detector stations spread over 17 km$^2$, detecting radio waves in the frequency range of $30-80$ MHz. AERA utilizes two channels, each one measuring in a different polarization: East-West and North-South relative to magnetic North. Although primarily focused on cosmic ray detection, its periodically triggered traces\footnote{Periodically triggered traces refer to the read-out requests made by each data acquisition system for all active stations every 100 seconds.} collected from 2014 until 2023 can be used to search for traces of solar activity.

The Sun, as our closest star, is an intense source of radio waves. The thermal radiation of the quiet Sun is further enhanced by intense radio bursts, during which energetic electrons stimulate high-frequency plasma waves, subsequently converting into electromagnetic radio waves. These emissions occur at the local electron plasma frequency, determined by the electron density of the plasma\cite{Melrose} given by

\begin{equation}
    f_{pe} = \frac{1}{2\pi} \sqrt{ \frac{N_e e^2}{m_e \epsilon_0} },
\end{equation}where $N_e$ is the electron number density of the plasma and $m_e$ the electron mass. Since $N_e$ decreases with height in the corona, there is a direct correlation between the detection of radio bursts on Earth and the height of their source within the solar corona. This makes the observation of Solar Radio Burst (SRB) in AERA data important in the context of studying the mechanism of Coronal Mass Ejection (CME) as the AERA frequency range corresponds to sources located in the upper corona. Furthermore, although in a frequency range distinct from AERA, increased levels of X-ray and extreme ultraviolet radiation during periods of heightened solar activity affect AERA data in two significant ways. Firstly, this radiation causes increased ionization of the upper ionosphere, leading to atmospheric reflections of terrestrial radio waves from distant sources. Secondly, higher ionization levels in the lower ionosphere result in the degradation or complete absorption (radio blackout) of radio signals.


The Sun follows an approximately 11-year cycle, with periods of maximum and minimum activity marked by variations in the number of sunspots and other indicators. During the phase of maximum activity, there is a substantial increase in the number of sunspots, which are regions of intense magnetic activity on the surface of the Sun. This peak of activity is associated with greater radiation emission and explosive events. On the other hand, the phase of minimum activity is characterized by a significant decrease in the occurrence of sunspots and lower magnetic activity.

\begin{figure}[H]
        \centering
                \includegraphics[scale=0.34]{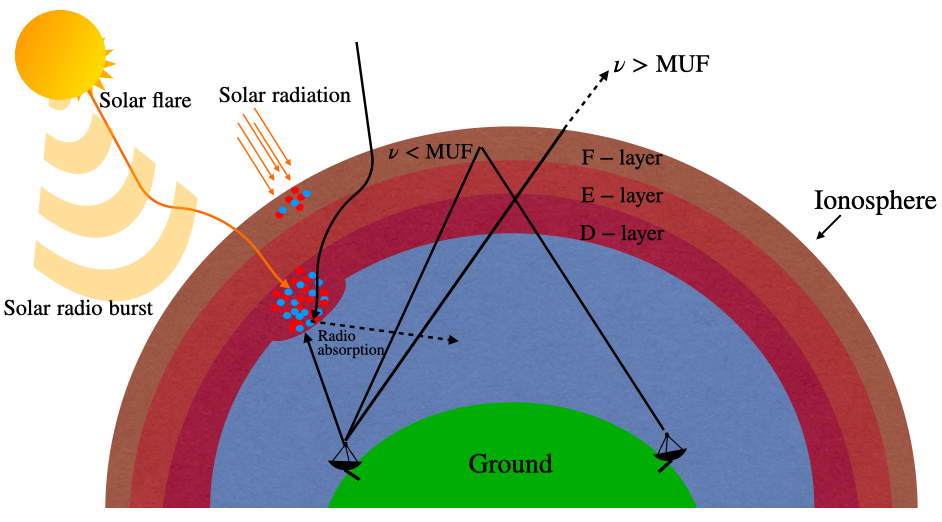}\quad
         \caption{\footnotesize{Schematic representation of the Earth's atmosphere, emphasizing the ionosphere with the F, E, and D layers. We can observe the propagation of a radio signal, emitted and reflected in the F layer of the ionosphere. The incidence of solar radiation is depicted, illustrating how it contributes to increased ionization of the F layer. The more intense ionization in this layer raises the MUF, causing the atmosphere to be transparent to radio signals with frequencies higher than the MUF or reflected back to Earth if the frequency is lower. Additionally, the dynamics of a solar flare emitted by the Sun are also depicted, highlighting the ionization of the innermost layer. This phenomenon can lead to a radio blackout as part of the signal is absorbed in the D layer. Finally, the presence of the solar burst in the figure emphasizes its contribution to a strong radio signal in the dynamic frequency spectrum.}}
         \label{Fig:ionospherschematic}
 \end{figure}

Figure \ref{Fig:ionospherschematic} shows a schematic representation of the Earth's atmosphere, illustrating the layers impacted by solar activity and how this influences the propagation of radio waves between 2 points located in the ground. Solar radiation, represented by orange arrows, plays a predominant role in modulating the ionospheric conditions, notably in the F layer. This phenomenon occurs due to the interaction of solar radiation, especially ultraviolet light, with the atoms and molecules present in these layers. During periods of high solar activity, the intensification of ultraviolet radiation results in a substantial increase in the ionization of these layers. As a result, atoms and molecules in this region lose electrons due to the absorption of this radiation, leading to a higher concentration of ions and free electrons. These variations in ionospheric ionization directly affect the Maximum Usable Frequency (MUF), which represents the highest frequency that can be used for radio communication between two points on Earth, taking ionospheric conditions into account. The MUF is mainly influenced by the electron density in the ionosphere. When the electron density is high, the MUF is also high, allowing higher frequencies to be used for long-distance radio communication as these waves are reflected by the ionosphere as shown in the figure. For frequencies above the MUF, the atmosphere becomes transparent, allowing the radio wave to pass through the F layer.

During a solar flare, radiation penetrates deeper into the atmosphere, reaching inner layers. The D layer, which normally does not contribute significantly to the reflection of radio waves, becomes a critical absorption point during the intense ionization caused by a solar flare. Radio waves that interact with electrons in this layer lose energy due to more frequent collisions, leading to considerable absorption. This phenomenon can result in a radio blackout, where high-frequency communication is impaired or completely absorbed, primarily in the 3 to 30 MHz range (as shown in Figure \ref{Fig:ionospherschematic}). In addition to impacts on radio communications, solar flares can also have effects in other areas, such as interference with satellite navigation systems.

Solar Radio Bursts (SRBs) are intense emissions of radio waves from the Sun, usually associated with solar eruptions. During a solar eruption, the sudden increase in energy release triggers the acceleration of charged particles within the solar magnetic field, generating radio waves at different frequencies. The occurrence of these eruptions is a necessary condition for an SRB. However, solar eruptions are not always accompanied by SRBs, and the occurrence of SRBs does not depend on the intensity of solar eruptions, as they can occur even in less intense eruptions. SRBs are classified into five types—Type I, II, III, IV, and V—based on their spectral characteristics\cite{TypeI,TypeII,TypeIII,TypeIV,TypeV,KUNDU}. These classifications reflect differences in frequency drift, duration, and association with solar phenomena like coronal mass ejections and solar flares.

The proceeding is organized as follows: Section \ref{section-muf} discusses the effect of the MUF, influenced by both the 11-year solar cycle and daily variations, on the broadband noise detected by AERA antennas in the 30–40 MHz range. Section \ref{section-SRB} presents observations of SRB signals in AERA data associated with 16 intense solar flares, recorded by National Oceanic and Atmospheric Administration (NOAA) when South America was sunlit, which caused moderate to strong radio blackouts. Finally, conclusions are provided in section \ref{conclusions}.

\section{Correlation between MUF and broadband noise measured by AERA within the 30-40 MHz frequency range}
\label{section-muf}

\begin{figure}[H]
        \centering
                \includegraphics[scale=0.44]{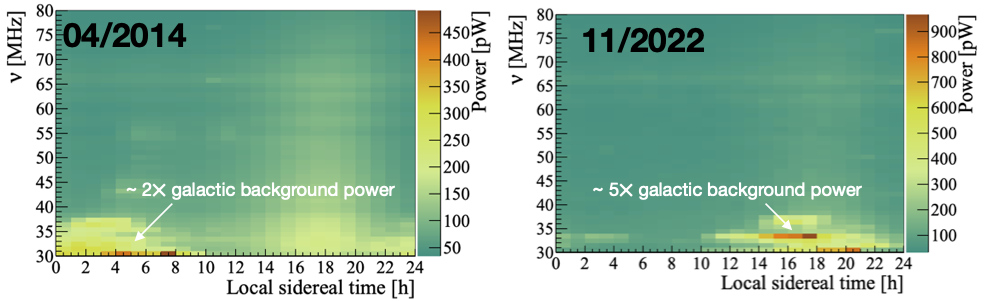}
    \caption{\footnotesize{Dynamic frequency spectrum collected in two distinct periods of time for one specific antenna (Id:33 antenna, east-west channel) highlighting the noise present in the 30-40 MHz frequency range.}}\label{Fig:spectrum_noises}
 \end{figure}

\hspace{0.5cm}The AERA data collected between 2014 and 2016 and after 2022 exhibit strong broadband noise in the 30-40 MHz band.  Figure \ref{Fig:spectrum_noises} depicts two distinct periods (March/2014 and September/2022) marked by exceptionally high average powers observed in the dynamic frequency spectrum within the specified frequency range. These noises are remarkably intense, being two to five times larger than the expected galactic background, effectively obscuring the modulation of the galactic background signal.

\begin{figure}[H]
        \centering
                \includegraphics[scale=0.30]{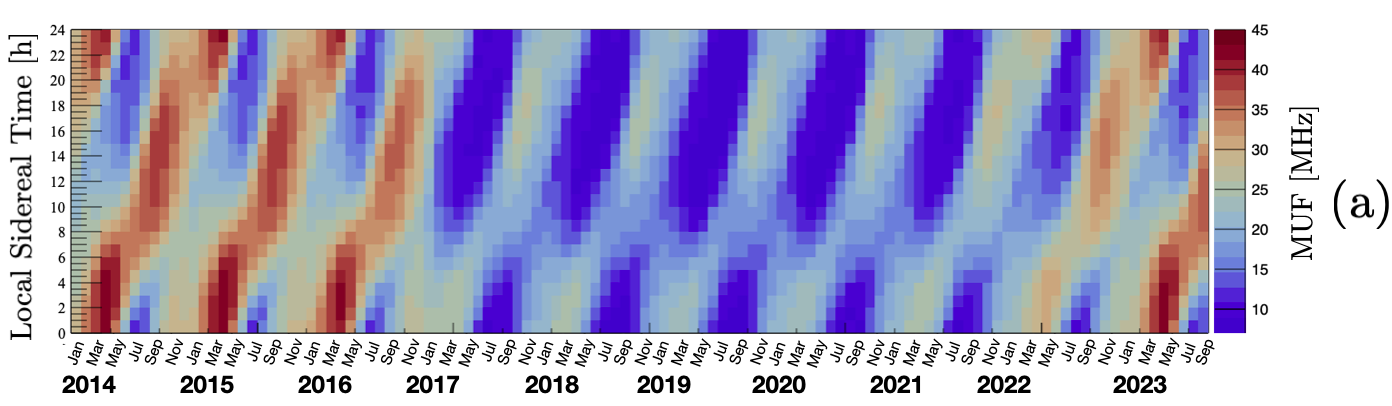}\quad
                \includegraphics[scale=0.30]{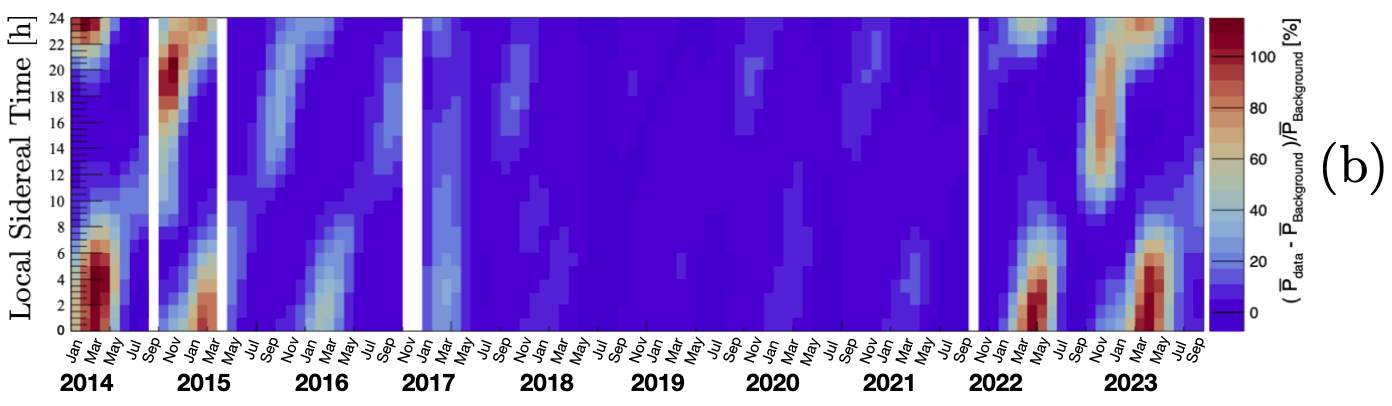}\quad
                \includegraphics[scale=0.30]{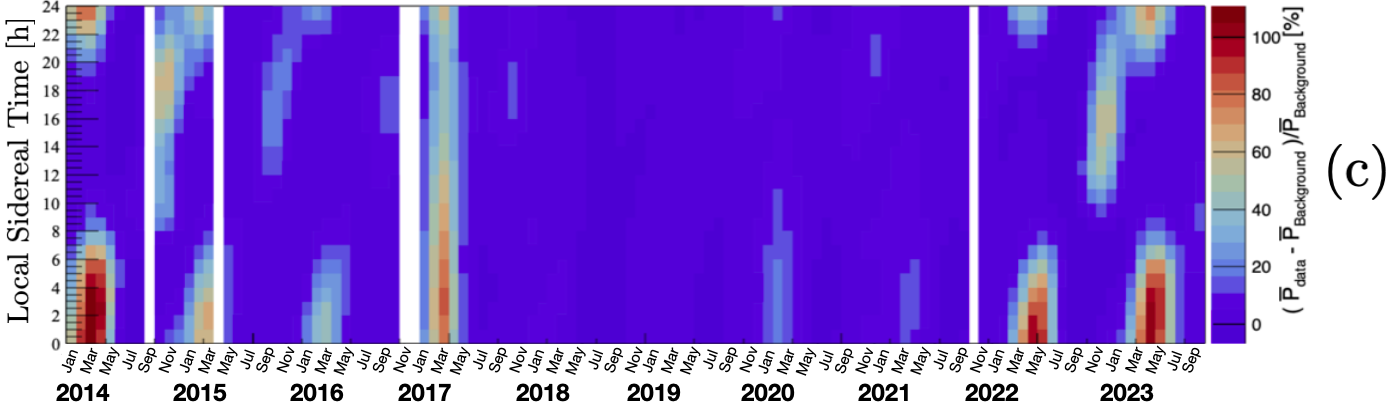}\quad
         \caption{\footnotesize{Comparison between the Maximum Usable Frequency and the power measured by the AERA stations. (a) Monthly MUF recorded in Malargüe region from 2014 to September 2023 as a function of local time and LST. (b) Relative power corresponding to 30-40 MHz measured in the same period of time for Butterfly antennas in the east-west channel. (c) the same as (b) for the north-south channel. White bands represent periods when no data is available. A clear correlation between the evolution of the MUF and the relative power is observed.}}\label{Fig:mufxtime}
 \end{figure}

These noises occur only during periods of high solar activity, which causes the MUF to increase due to the heightened ionization of the ionosphere. Normally, the MUF at Malargüe does not exceed 30 MHz (the beginning of AERA's frequency range). However, during periods of high solar activity, the MUF can surpass this threshold, allowing radio waves from distant sources to be reflected by the ionosphere and impinge on the Auger Observatory. To illustrate this correlation, Figure \ref{Fig:mufxtime} shows in panel (a) the MUF recorded on an hourly basis\footnote{The Maximum Usable Frequency is monitored locally through various stations around the world. The IRI software, available at \url{https://kauai.ccmc.gsfc.nasa.gov/instantrun/iri/}, compiles this data and interpolates to obtain the MUF in any region of the world. In this study, we utilize the MUF in the Malargüe region with coordinates: Latitude: -35.47, Longitude: -69.58.} from 2014 to September 2023. Panels (b) and (c) show the percentage differences of the data $\bar{P}_{\rm{data}}(t)$ relative to the galactic background $\bar{P}_{\rm{background}}(t)$ over the same period for the East-West and North-South channels, respectively. The plot clearly demonstrates that elevated MUF levels, caused by increased solar activity, directly correlate with the observed broadband noise in AERA data. This strongly suggests that the observed noises directly stem from the heightened solar activity, leading to increased ionization in the upper layers of the ionosphere which, in turn, facilitates the atmospheric reflection of radio waves emitted from distant sources on Earth.

\section{Detection of SRBs with AERA Data} 
\label{section-SRB}

\hspace{0.5cm} Solar flares are large events that release large amounts of radiation and particles into the interstellar medium. When this flare is directed toward the Earth, these events have the potential to modify the radioelectric environment and affect the propagation of radio waves. Furthermore, solar flares can initiate SRBs, which are intense radio emissions from the Sun that can leave their signatures on the data collected by AERA. However, detecting these events in AERA data is challenging, as specific conditions must align for these radio signals to be identified by the antennas.

One of the necessary conditions is that, at the time of the event, South America, where the Pierre Auger Observatory is located, must be illuminated by the sun. This is crucial because we are interested in the electromagnetic radiation released by the flare, which propagates at the speed of light. Consequently, the events are detected only 8 minutes after the flare occurs.  These energetic events are rare and occur more frequently at the peak of solar activity, which last happened in 2014 and is happening again in 2024. Furthermore, as a consequence of large flares, another phenomenon can occur, which is radio blackouts\footnote{Large solar flares, with possible radio blackouts, are monitored and recorded by the NOAA \url {https://www.swpc.noaa.gov/}} as described in section \ref{sec:intro}. However, these blackouts are more common in the band between 3-30 MHz, and are rare at the AERA higher frequencies, since a very intense solar flare is required.

 \begin{table}[H]
 \centering
\footnotesize
\begin{tabular}{|c|c|c|c|c|c|c|c|}
\hline
\textbf{N$^{\circ}$} & \textbf{Date} & \textbf{\begin{tabular}[c]{@{}c@{}}UTC\\ Start\end{tabular}} & \textbf{\begin{tabular}[c]{@{}c@{}}UTC\\ Maximum\end{tabular}} & \textbf{\begin{tabular}[c]{@{}c@{}}UTC\\ End\end{tabular}} & \textbf{\begin{tabular}[c]{@{}c@{}}Maximum\\  flux\end{tabular}} & \textbf{\begin{tabular}[c]{@{}c@{}}Radio\\  Blackout\end{tabular}} & \textbf{\begin{tabular}[c]{@{}c@{}}Severity\\  Descriptor\end{tabular}} \\ \hline
\textbf{1}           & 2014/09/10    & 17:21                                                        & 17:45                                                          & 18:20                                                      & X2.39                                                            & R3                                                                 & Strong                                                                  \\ \hline
\textbf{2}           & 2014/10/02    & 18:49                                                        & 19:01                                                          & 19:14                                                      & X1.05                                                            & R3                                                                 & Strong                                                                  \\ \hline
\textbf{3}           & 2014/10/22    & 14:02                                                        & 14:28                                                          & 14:50                                                      & X2.39                                                            & R3                                                                 & Strong                                                                  \\ \hline
\textbf{4}           & 2014/10/27    & 14:12                                                        & 14:47                                                          & 15:09                                                      & X2.96                                                            & R3                                                                 & Strong                                                                  \\ \hline
\textbf{5}           & 2017/09/10    & 15:35                                                        & 16:06                                                          & 16:31                                                      & X11.88                                                           & R3                                                                 & Strong                                                                  \\ \hline
\textbf{6}           & 2021/07/03    & 14:18                                                        & 14:28                                                          & 14:34                                                      & X1.59                                                            & R3                                                                 & Strong                                                                  \\ \hline
\textbf{7}           & 2021/10/28    & 15:17                                                        & 15:35                                                          & 15:48                                                      & X1.0                                                             & R3                                                                 & Strong                                                                  \\ \hline
\textbf{8}          & 2022/03/31    & 18:17                                                        & 18:35                                                          & 18:45                                                      & M9.67                                                            & R2                                                                 & Moderate                                                                \\ \hline
\textbf{9}          & 2023/02/11    & 15:40                                                        & 15:48                                                          & 15:54                                                      & X1.1                                                             & R3                                                                 & Strong                                                                  \\ \hline
\textbf{10}          & 2023/02/28    & 17:35                                                        & 17:50                                                          & 17:56                                                      & M8.62                                                            & R2                                                                 & Moderate                                                                \\ \hline
\textbf{11}          & 2023/03/03    & 17:42                                                        & 17:52                                                          & 17:59                                                      & X2.07                                                            & R3                                                                 & Strong                                                                  \\ \hline
\textbf{12}          & 2023/11/28    & 19:35                                                        & 19:50                                                          & 20:09                                                      & M9.82                                                            & R2                                                                 & Moderate                                                                  \\ \hline
\textbf{13}          & 2023/12/14    & 16:47                                                        & 17:02                                                          & 17:12                                                      & X2.87                                                            & R3                                                                 & Strong                                                                \\ \hline
\textbf{14}          & 2024/05/09    & 17:23                                                        & 17:44                                                          & 18:01                                                      & X1.1                                                            & R3                                                                 & Strong                                                                \\ \hline
\textbf{15}          & 2024/05/14    & 16:46                                                        & 16:51                                                          & 17:02                                                      & X8.79                                                            & R3                                                                 & Strong                                                                \\ \hline
\textbf{16}          & 2024/05/15    & 14:20                                                        & 14:38                                                          & 14:51                                                      & X2.9                                                            & R3                                                                 & Strong                                                                \\ \hline
\end{tabular}\caption{\footnotesize{Record of solar flare events with the potential to cause radio blackout between 2014 and 2023, when South America was sunlit. The table is organized chronologically and provides information about the beginning and end of each event, along with the maximum flux reached. It also includes details about the type of radio blackout associated with each event, along with a description of the severity of the impact.}}\label{Table_events}
\end{table}

Solar flares are categorized by X-ray intensity into classes X (highest intensity), M (moderate), C, and B (lowest intensity)\cite{GOES_NOAA}. Radio blackouts, associated with Class X and M flares, are rated by NOAA from R1 to R5, indicating severity from minor to extreme \cite{NOAA_scale}. For instance, an R3 blackout can significantly disrupt high-frequency communications, while an R5 blackout can cause complete communication failure.  We investigated 16 events reported by NOAA from 2014 to 2024, which had the potential to cause significant disturbances in radio waves and temporary interruptions due to R2 and R3 blackouts. These events occurred when South America was sunlit and included flares of Class X and M intensities. Table \ref{Table_events} provides a chronological list of these events. While all events left significant signals in the AERA data, we will present only two of them as examples: Event N$^{\circ}$7 class X1.0 and Event N$^{\circ}$3 class X2.39.

   \begin{figure}[H]
        \centering
                \includegraphics[scale=0.36]{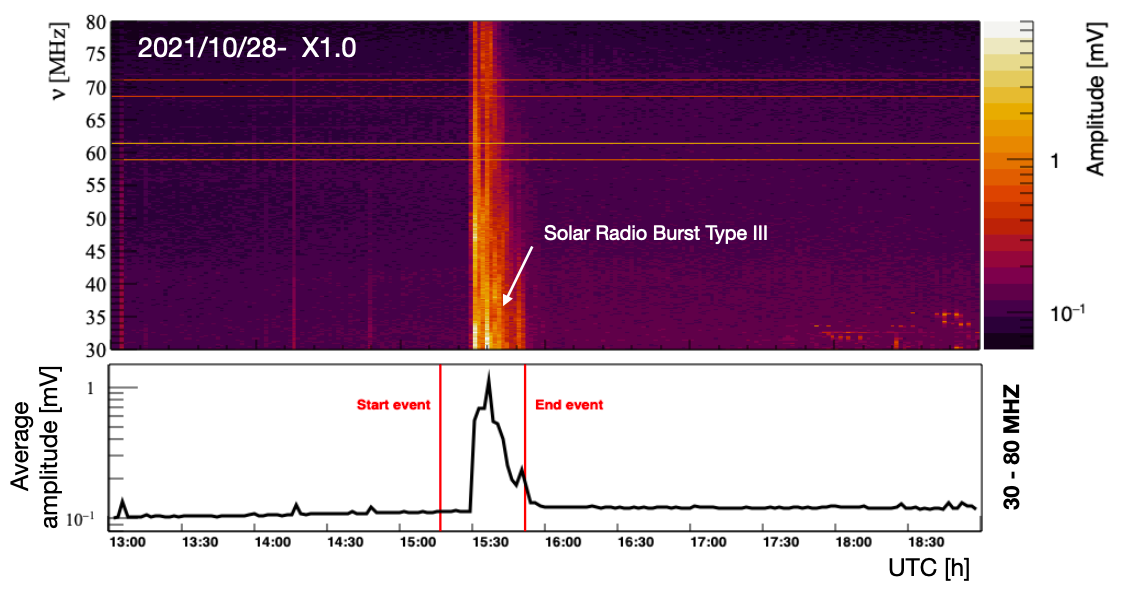}\quad
         \caption{\footnotesize{Average dynamic frequency spectrum as a function of UTC, collected by all Butterfly antennas around the time of Event N$^{\circ}$7, classified as X1.0, on October 28, 2021.  The bottom panel displays the signal average across all frequencies from 30 to 80 MHz. Vertical red lines indicate the start and end of the solar event.}}\label{Fig:event_n7}
 \end{figure}

     \begin{figure}[H]
        \centering
                \includegraphics[scale=0.36]{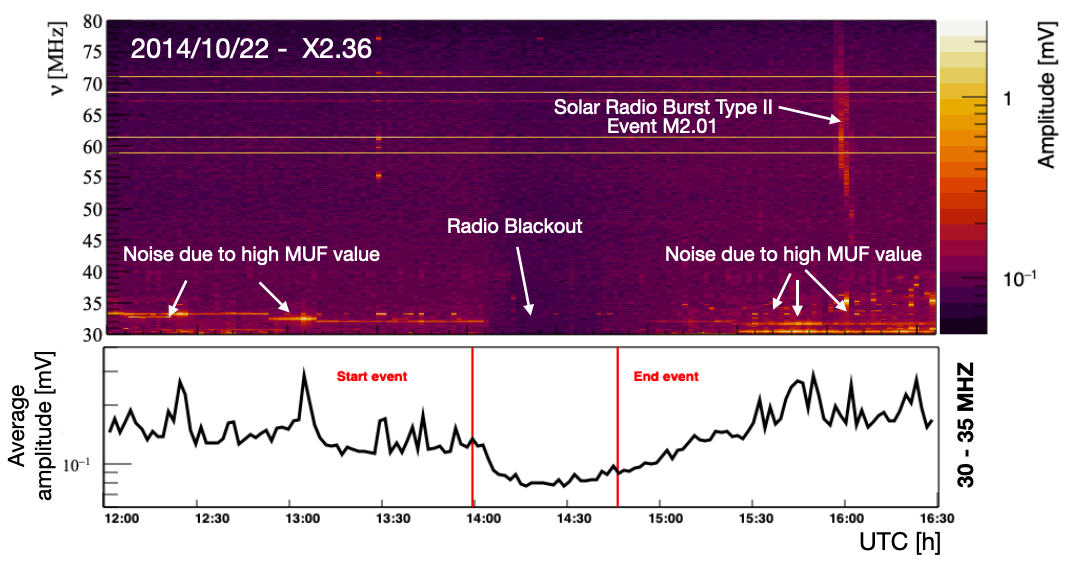}\quad
         \caption{\footnotesize{Average dynamic frequency spectrum as a function of UTC, captured by all Butterfly antennas around the time of Event N$^{\circ}$3 classified as X2.36, on October 22, 2014. The spectrogram illustrates a radio blackout occurrence during the event, along with additional noise attributed to the high MUF value. The bottom panel displays the signal average across all frequencies from 30 to 35 MHz. Vertical red lines indicate the start and end of the solar event.}}\label{Fig:event_n3}
 \end{figure}

Event N$^{\circ}$7, with an intensity of X1.0, triggered an SRB and lasted for 31 minutes. It left a clear signature in the average dynamic frequency spectrum measured by AERA, covering the entire frequency range. Figure \ref{Fig:event_n7} shows this in a $\sim2$ hours window around the event, highlighting the type III SRB characteristic signature during the flare. The bottom panel shows the average signal across the frequency range, illustrating the event's peak.

The event N$^{\circ}$3 caused a radio blackout in the AERA data, as shown in Figure \ref{Fig:event_n3}. The impact is observed as a decrease in the amplitude of the AERA data precisely during the event's duration, indicated in the top panels of the figure. Additionally, this event produced a radio blackout on frequencies below 30 MHz used in telecommunications, reported by NOAA. The same dynamic frequency spectra reveal large power values around 30 MHz, resulting from high MUF values, characteristic of the peak solar activity in 2014. The bottom panels of the figure show the average signal integrated in the 30-35 MHz band, where the radio blackout had a pronounced influence, clearly attenuating the power due to increased ionization in the lower ionosphere layers. It is important to highlight that approximately 2 hours after event N$^{\circ}$ 3, another event of M2.01 intensity was detected. This event, despite not having been previously alerted by NOAA as a potential radio blackout due to its low intensity, is clearly visible in the data collected by AERA with a signature of a type II solar radio burst.
 
\section{Conclusions}
\label{conclusions}
We analyzed the impact of solar activity on AERA data collected over a decade. Our study revealed a strong correlation between the MUF, modulated by the solar cycle, and broadband noise in the 30-40 MHz range. This correlation indicates that the observed noise likely results from increased solar activity, leading to greater ionization in the upper ionosphere and enhancing the atmospheric reflection of terrestrial radio waves emitted from distant sources on Earth. Additionally, we identified signals from 16 Solar Radio Bursts in the AERA data, associated with moderate to strong radio blackouts, particularly relevant for understanding coronal mass ejections, as AERA's frequency range (30-80 MHz) captures emissions from the upper solar corona. Although AERA was originally designed for detecting radio waves from extensive air showers, our findings reveal its broader potential for studying solar activity and its effects on Earth's atmosphere. This opens new possibilities for exploring various solar and atmospheric phenomena, particularly those related to solar activity and its interactions with the Earth's atmosphere.


\end{document}